# Investigation of the energy-averaged double transition density of isoscalar monopole excitations in medium-heavy mass spherical nuclei

M.L. Gorelik[1,2], S. Shlomo[1,2], B.A. Tulupov[1,3], and M.H. Urin[1]

[1]*National Research Nuclear University «MEPhI», Moscow, 115409 Russia*
[2]*Cyclotron Institute, Texas A&M University, College Station, Texas 77843, USA*
[3]*Institute for Nuclear Research, RAS, Moscow, 117312 Russia*

The particle-hole dispersive optical model, developed recently, is applied to describe properties of high-energy isoscalar monopole excitations in medium-heavy mass spherical nuclei. We consider, in particular, the double transition density averaged over the energy of the isoscalar monopole excitations in $^{208}$Pb in a wide energy interval, which includes the isoscalar giant monopole resonance and its overtone. The energy-averaged strength functions of these resonances are also analyzed. Possibilities for using the mentioned transition density to description of inelastic α-scattering are discussed.

## I. INTRODUCTION

The study of properties of collective states in nuclei provides information on the bulk properties of nuclear matter. In particular, the interest in experimental and theoretical studies of high-energy particle-hole-type isoscalar monopole (ISM) excitations in medium-heavy mass nuclei is mainly due to the possibility of determining the nuclear matter incompressibility coefficient, a fundamental physical quantity essential for astrophysics and nuclear physics. The value of this coefficient depends on the mean energy of the strength distribution, corresponding to the ISM external field $r^2 Y_{00}$ (in other words, on the energy of the isoscalar giant monopole resonance (ISGMR)) [1]. To deduce this strength distribution from experimental data of (α,α')-inelastic scattering cross sections at small angles, it is usually assumed that the ISM strength is concentrated in the vicinity of the ISGMR and the properly normalized classical collective model transition density of the ISGMR can be used within the folding model distorted wave Born approximation (FM-DWBA) (see. e.g., Ref. [2] for details). It is important to point out that the classical collective model transition density is independent of the excitation energy. In Ref. [3], a microscopic evaluation of the (α,α')-scattering cross sections were carried out within the FM-DWBA using the Hartree-Fock (HF) ground state density and the transition densities obtained from the HF-based random-phase-approximation (RPA) calculations. A comparison with results obtained with the classical collective model transition density was also made in Ref. [3].

The main properties (including direct-nucleon-decay probabilities) of various isoscalar giant resonances were investigated in details in Refs. [4,5] within the semi-microscopic approach

(see reviews in Refs. [6,7]). However, the validity of this approach is limited to a vicinity of the given giant resonance. The following points stimulated us to carry out the theoretical study presented in this paper: (i) the unexpected results of Ref. [2] concerning the observed strength of a high-energy component of the ISM excitations in the A≈90 region; (ii) the necessity to examine from a microscopic point of view the applicability of the classical collective model transition densities of the ISGMR and its overtone (ISGMR2) in the analysis of the corresponding experimental data.

We emphasize that in a microscopic approach, the input quantity for the analysis of the (α,α')-reaction cross section should be the energy-averaged double transition density (i.e. the energy-averaged product of energy dependent transition densities taken in different points). In a wide excitation-energy interval involving the ISGMR and its overtone, ISGMR2, this quantity is expected to be different from the product of the classical collective model transition densities [2], which is independent of excitation energy, or the product of microscopic transition densities [3], due to proper treatment of the shell structure of nuclei (i.e. the Landau damping) and also the spreading effect. The particle-hole (p-h) dispersive optical model (PHDOM), developed recently (see Refs. [7,8]), allows one to describe the energy-averaged ISM double transition density at arbitrary (but high-enough) excitation energy and, in particular, to trace the change of this quantity from the ISGMR to ISGMR2.

The PHDOM, as an extension of the continuum-RPA, accounts for the Landau damping, coupling of high-energy (p-h)-type states to the single-particle (s-p) continuum and to many-quasiparticle configurations (the spreading effect). Within the PHDOM, the Landau damping and coupling of high-energy (p-h)-type states to the s-p continuum are described microscopically in terms of the Landau-Migdal p-h interaction and a phenomenological mean field, partially consistent with this interaction. The spreading effect is treated phenomenologically and with averaging over the energy in terms of the imaginary part of an effective s-p optical-model potential. The imaginary part also determines the corresponding real part, added to mean field, via the proper dispersive relationship.

In Ref. [9] we have presented the first evaluation of the energy-averaged double p-h transition density within the PHDOM with very limited results. In this work we present results of extended investigation of the energy-averaged double transition density. The calculations are performed for ISM excitations in $^{208}$Pb. A wide excitation-energy interval is considered which includes the ISGMR and ISGMR2. The fractions of the energy-weighted sum-rule (EWSR) associated with the strength functions of these resonances, i.e. the energy-weighted strength functions divided by the corresponding EWSR, are analyzed. We also consider single p-h transition density obtained by various projections of the double p-h transition density, as well as the classical collective models transition densities, and investigate their applicability. This is done by considering the excitation cross sections with the Born approximation and comparison with the results obtained from the double p-h transition density.

## II. FORMALISM

Within the PHDOM, described in Refs. [7,8], the continuum-RPA of Ref. [10] is extended to take into account the spreading effect phenomenologically with averaging over the energy. The basic ingredient of the model is the energy-averaged p-h Green function (or the effective p-h propagator), $A^{\alpha\beta}$, where $\alpha,\beta = n,p$ are the isobaric indexes. The "free" p-h propagator, which corresponds to the model of independent and damping quasiparticles, is diagonal in these indexes: $A_0^{\alpha\beta} = A_0^\alpha \delta_{\alpha\beta}$. The properties of isoscalar excitations are described by a propagator obtained as a proper sum: $A = \sum_{\alpha,\beta=n,p} A^{\alpha\beta}$, with $A_0 = \sum_{\alpha=n,p} A_0^\alpha$. The ISM radial component of this propagator, $(r^2 r'^2)^{-1} A(r,r',\omega)$, determines the energy-averaged ISM radial double transition density

$$-\frac{1}{\pi}\mathrm{Im}\, A(r,r',\omega) = \langle \rho(r,\omega)\rho(r',\omega) \rangle \equiv \rho(r,r',\omega). \qquad (1)$$

In Eq. (1), $\omega$ is the excitation energy ($\omega > B_\alpha$, $B_\alpha$ is the nucleon separation energy), $\rho(r,\omega) = \sum_{\alpha=n,p} \rho^\alpha(r,\omega)$ is the radial one-dimensional transition density and the brackets $\langle ... \rangle$ denote averaging over the energy. We point out that the three-dimensional ISM transition density is determined as $\rho(\vec{r},\omega) = r^{-2}\rho(r,\omega)Y_{00}(\vec{n})$ so that the corresponding ISM double transition density is given by:

$$\rho(\vec{r},\vec{r}',\omega) = \frac{1}{4\pi}(rr')^{-2}\rho(r,r',\omega). \qquad (2)$$

In accordance with Eq. (1), the energy-averaged strength function $S_{V_0}(\omega)$, associated with the ISM external s-p field $V_0(\vec{r}) = V_0(r)Y_{00}(\vec{n})$, is determined by the averaged effective p-h propagator:

$$S_{V_0}(\omega) = -\frac{1}{\pi}\mathrm{Im}\int V_0(r) A(r,r',\omega) V_0(r') dr dr' = -\frac{1}{\pi}\mathrm{Im}\int V_0(r) A_0(r,r',\omega) V(r',\omega) dr dr'. \qquad (3)$$

We add that within the PHDOM, there is an equivalent and more simple method for evaluation of the energy-averaged strength function by using the effective field $V(r,\omega)$, which corresponds to an external s-p field $V_0(r)$. The effective field, determined by the equation

$$\int A(r,r',\omega) V_0(r') dr' = \int A_0(r,r',\omega) V(r',\omega) dr', \qquad (4)$$

satisfies a more simple equation than the below-given equation for $A(r,r',\omega)$ (see, e.g., Ref. [4-6]). However, this method cannot be used for the evaluation of the energy-averaged double transition density [7,8]. To describe the properties of the ISGMR and ISGMR2, it is convenient to choose for the radial external fields the forms

$$V_{0,1}(r) = r^2, \quad V_{0,2}(r) = r^4 - \eta r^2, \qquad (5)$$

respectively. In the present work, unlike the approaches of Refs. [5,11], we find the parameter $\eta$ by minimizing the quantity $x_2(\eta) = \int y_2(\omega) d\omega$, $y_2(\omega) = \omega S_{V_{0,2}}(\omega)/(EWSR)_{V_{0,2}}$, where the integration is performed over the ISGMR region.

We point out that the Bethe-Goldstone equation for the energy-averaged effective p-h propagator is given in Refs. [7,8] in a rather general form. This equation is given below in the form which is directly used for the evaluation of the ISM basic quantity $((N-Z) \ll A)$:

$$A(r,r',\omega) = A_0(r,r',\omega) + \int A_0(r,r_1,\omega) F(r_1) A(r_1,r',\omega) r_1^{-2} dr_1 . \tag{6}$$

In (6), $F(r)$ is the strength of the isoscalar part of the Landau-Migdal p-h interaction [12]: $F(\vec{r}_1,\vec{r}_2) \to F(r_1)\delta(\vec{r}_1 - \vec{r}_2)$. In accordance with Refs. [7,8,13], the expression for the "free" p-h propagator $A_0(r,r',\omega)$ can be presented in the form (the isobaric index $\alpha$ is omitted for brevity):

$$A_0(r,r',\omega) = \sum_{\mu,(\lambda)} n_\mu t^2_{(\lambda)(\mu)} \chi_\mu(r)\chi_\mu(r') g_{(\lambda)}(r,r',\varepsilon_\mu + \omega) + \sum_{\lambda,(\mu)} n_\lambda t^2_{(\mu)(\lambda)} \chi_\lambda(r)\chi_\lambda(r') g_{(\mu)}(r,r',\varepsilon_\lambda - \omega) +$$
$$+ 2\sum_{\mu,(\lambda)} n_\mu n_\lambda t^2_{(\mu)(\lambda)} \chi_\mu(r)\chi_\mu(r')\chi_\lambda(r)\chi_\lambda(r') \frac{(iW(\omega) - P(\omega)) f_\mu f_\lambda}{(\varepsilon_\lambda - \varepsilon_\mu - \omega)^2 + (iW(\omega) - P(\omega))^2 f_\mu^2 f_\lambda^2} . \tag{7}$$

In (7), $t_{(\lambda)(\mu)} = \frac{1}{4\pi}(2j_\mu + 1)^{1/2}\delta_{(\lambda)(\mu)}$, $n_\mu$ is the occupation factor for the s-p level $\mu$, which is characterized by the energy $\varepsilon_\mu$ and by the set of quantum numbers of the total and orbital momenta $(\mu) \equiv (j_\mu, l_\mu)$; $r^{-1}\chi_\mu(r)$ is the bound state radial wave function, $f_\mu = \int f_{WS}(r) \chi_\mu^2(r) dr$ and $f_{WS}(r)$ is the well-known Woods-Saxon function. Being the imaginary and real parts of the strength of an energy-averaged specific p-h interaction responsible for the spreading effect, the phenomenological quantities $W(\omega)$ and $P(\omega)$ determine the optical-model-like addition to the nuclear mean field used in the calculations of the optical-model radial Green functions $(rr')^{-1} g(r,r',\omega)$:

$$\left(h_{0,(\lambda)}(r) - \left\{\varepsilon_\mu + \omega + (iW(\omega) - P(\omega)) f_\mu f_{WS}(r)\right\}\right) g_{(\lambda)}(r,r',\varepsilon_\mu + \omega) = -\delta(r-r'), \tag{8}$$

$$\left(h_{0,(\mu)}(r) - \left\{\varepsilon_\lambda - \omega + (iW(\omega) - P(\omega)) f_\lambda f_{WS}(r)\right\}\right) g_{(\mu)}(r,r',\varepsilon_\lambda - \omega) = -\delta(r-r'). \tag{9}$$

Here, $(\lambda) = (\mu)$ and $h_0(r)$ are the radial parts of the s-p Hamiltonian (including the spin-orbit and centrifugal terms).

Completing the brief description of the basic PHDOM equations, we note that there is a weak violation of the model unitarity. The violation is caused by the energy dependence of the effective optical-model potential and also by the use of the approximate spectral expansion for the optical-model Green functions of Eqs. (8) and (9). Methods for restoration of the model unitarity are under consideration. Some preliminary results are presented in Ref. [14].

It is common to use the energy-averaged classical double transition densities of the ISGMR and ISGMR2, corresponding to single-level resonances and presented in the factorized form:

$$\rho^L_{c,i}(r,r',\omega) = \rho^L_{c,i}(r,\omega)\rho^L_{c,i}(r',\omega) = L_i(\omega)\rho_{c,i}(r)\rho_{c,i}(r') , \tag{10}$$

$$L_i(\omega) = \frac{\Gamma_i}{2\pi}\left((\omega - \omega_i)^2 + \frac{1}{4}\Gamma_i^2\right)^{-1} .$$

Here, $\omega_i$ and $\Gamma_i$ are the resonance energy and total width, respectively. The classical collective model transition densities for the ISGMR and ISGMR2, normalized to the corresponding energy-weighted sum rules, are given by [11]:

$$\rho_{c,1}(r) = C_1 r^2 \sqrt{4\pi}\left(3 + r\frac{d}{dr}\right)n(r), \qquad C_1^2 = \frac{\hbar^2}{2mA\omega_1\langle r^2\rangle}, \tag{11}$$

$$\rho_{c,2}(r) = C_2 r^2 \sqrt{4\pi}\left(10r^2 - 3\eta_c + r(2r^2 - \eta_c)\frac{d}{dr}\right)n(r), \tag{12}$$

$$C_2^2 = \frac{\hbar^2}{2mA\omega_2\left(4\langle r^6\rangle - 4\eta_c\langle r^4\rangle + \eta_c^2\langle r^2\rangle\right)}, \quad \eta_c = 2\langle r^4\rangle/\langle r^2\rangle,$$

where $n(r) = \sum_\alpha n^\alpha(r)$ is the ground state matter density, $m$ is the nucleon mass, the brackets $\langle...\rangle$ mean here averaging over the mentioned density. We point out that in the present work we do not adopt the strength of the classical transition densities defined via the Lorentzian $L_i(\omega)$, Eq. (10). To conform with the widely-used method of analyzing the experimental excitation cross sections, we instead normalize the classical transition density $\rho_{c,i}^\Lambda(r,\omega) = \Lambda_i^{1/2}(\omega)\rho_{c,i}(r)$ by imposing the condition of reproducing the microscopic energy-weighted strength functions, i.e. imposing the condition

$$\left(\int \rho_{c,i}^\Lambda(r,\omega)V_{0,i}(r)dr\right)^2 = \int V_{0,i}(r)\rho(r,r',\omega)V_{0,i}(r')drdr', \tag{13}$$

using Eqs. (5), (11) and (12) for the external fields, the classical transition densities of the ISGMR and ISGMR2. As follows from Eqs. (1), (3) and (13), $\Lambda_i(\omega) = S_{V_{0,i}}(\omega)/(\rho_{c,i}(r)V_{0,i}(r)dr)^2 = \omega_i S_{V_{0,i}}(\omega)/(EWSR)_{V_{0,i}}$. The resulting microscopically corrected classical double transition densities, $\rho_{c,i}^\Lambda(r,r',\omega)$, will be compared with that obtained in our microscopic calculations.

We also study possibilities of an appropriate factorization of the ISM double-transition-density radial dependence in a wide excitation-energy interval. For this purpose we define the projected transition density (adopted in Ref. [3]),

$$\rho_{V_0}(r,\omega) = \int \rho(r,r',\omega)V_0(r')dr'\Big/S_{V_0}^{1/2}(\omega), \tag{14}$$

which can be considered as the transition density of a given ISM giant resonance (ISGMR or its overtone). This transition density, employed in Refs. [3-6,11] as the energy-dependent transition density of the corresponding giant resonance, fulfils the condition that the strength function is obtained by

$$S_{V_0}(\omega) = \left(\int \rho_{V_0}(r,\omega)V_0(r)dr\right)^2. \tag{15}$$

The projected transition density can be expressed also in terms of the effective field [4-6]:

$$\rho_{V_0}(r,\omega) = -\frac{1}{\pi}\mathrm{Im}\frac{V(r,\omega)}{F(r)S_{V_0}^{1/2}(\omega)}. \tag{16}$$

The distorted wave Born approximation (DWBA) is widely used in the analysis of measured cross sections of scattering probes. The energy-averaged DWBA differential cross section for the excitation of a giant resonance by inelastic α-scattering is given by

$$\left\langle\frac{d\sigma}{d\Omega_{\alpha'}dE_{\alpha'}}^{DWBA}\right\rangle = \left(\frac{\mu}{2\pi\hbar^2}\right)^2\frac{k_{\alpha'}}{k_\alpha}\left\langle\sum_s|T_{\alpha's\alpha 0}|^2\delta(E_s - E_0 - \omega)\right\rangle, \tag{17}$$

where, $\mu$ is the α-particle reduced mass and $\hbar k_\alpha = \sqrt{2\mu E_\alpha}$ and $\hbar k_{\alpha'} = \sqrt{2\mu(E_\alpha - \omega)}$ are the initial and final momenta of the α-nucleus relative motion, respectively. The transition matrix element is given by

$$T_{\alpha's\alpha 0} = \left\langle \varphi_{\alpha'}^{(-)}(\vec{r}_\alpha)\psi_s(\vec{r}_i) \middle| V_{\alpha A} \middle| \varphi_\alpha^{(+)}(\vec{r}_\alpha)\psi_0(\vec{r}_i) \right\rangle. \tag{18}$$

Here, $V_{\alpha A} = \sum_i V(|\vec{r}_\alpha - \vec{r}_i|)$ is the α-nucleus interaction, where $V$ is the α-nucleon interaction and $\vec{r}_\alpha$ and $\vec{r}_i$ are the coordinates of the α-particle and the nucleon, respectively. In Eq. (18), $\psi_0$ and $\psi_s$ are the wave functions of the ground and excited states of the nucleus, and $\varphi_\alpha^{(+)}$, $\varphi_{\alpha'}^{(-)}$ are the corresponding distorted wave functions of the α-nucleus relative motion. We emphasize that the energy-averaged cross-section is obtained from Eq. (17) by using the energy-averaged transition strength function. Therefore, the input to the DWBA calculations should be the energy-averaged double transition density $\rho(r,r',\omega)$ of Eq. (1) (see the spectral expansion of $A(r,r',\omega)$, e.g., in Refs. [7,8]).

In the experimental analysis of inelastic (α,α')-scattering cross sections at small angles the folding model (FM)-DWBA is usually employed. One first determines the optical potential by folding a Fermi distribution for the ground state matter density with a parameterized α-nucleon interaction, determined by a fit to the elastic scattering cross-section. Next, the classical collective ISGMR and ISGMR2 transition densities of Eqs. (11) and (12) are folded with the α-nucleon interaction to determine the transition potentials. Then the optical potential and the transition potentials are used as input for the DWBA code (see, e.g., Refs. [2,3]). Since only the energy-averaged ISM double transition density of Eq. (1) can be obtained within the PHDOM, we study the accuracy of this approach by calculating the excitation cross section within the Born approximation. In this approximation, the energy averaged transition strength function is proportional to the strength function $S_{V_{0,q}}$, corresponding to the external field $V_{0,q}(r) = \dfrac{\sin(qr)}{qr}$,

$$\left\langle \sum_s |T_{\alpha's\alpha 0}|^2 \delta(E_s - E_0 - \omega) \right\rangle = |V_{\alpha A}(q)|^2 \int V_{0,q}(r)\rho(r,r',\omega)V_{0,q}(r')drdr'. \tag{19}$$

Here, $V_{\alpha A}(q)$ is the Fourier transform of the α-particle-nucleon interaction, $q = k_\alpha - k_{\alpha'}$, with $E_\alpha$ related to the projectile energy of the α-particle, used in the experiments of Ref. [2]). We note that since the classical double transition densities for the ISGMR and ISGMR2 are factorized the cross-sections can be directly obtained from the transition strength function using the transition densities $\rho_{c,i}^\Lambda(r,r',\omega)$ of Eq. (13).

To check the possibility of the approximate factorization of the ISM double-transition-density $\rho(r,r',\omega) \approx \rho_{V_{0,i}}(r,\omega)\rho_{V_{0,i}}(r',\omega)$ in the energy regions of the ISGMR and the ISGMR2 we examine the quantity

$$Q_{V_{0,i}}(\omega) = \dfrac{\left(\int \rho_{V_{0,i}}(r,\omega)V_{0,q}(r)dr\right)^2}{\int V_{0,q}(r)\rho(r,r',\omega)V_{0,q}(r')drdr'}, \tag{20}$$

where, the definition of Eqs. (14) and (15) is used for $\rho_{V_{0,i}}(r,\omega)$. For the same aim the microscopically corrected classical double transition densities of Eq. (13) can be also used:

$$Q_{c,i}^{\Lambda}(\omega) = \frac{\int V_{0,q}(r)\rho_{c,i}^{\Lambda}(r,r',\omega)V_{0,q}(r')drdr'}{\int V_{0,q}(r)\rho(r,r',\omega)V_{0,q}(r')drdr'}. \tag{21}$$

### III. INPUT QUANTITIES AND RESULS OF CALCULATIONS

The input quantities of the PHDOM calculations carried out in this work are the Landau-Migdal p-h interaction and a partially self-consistent phenomenological mean field together with the imaginary part of the effective optical-model potential. The parameters of the mean field consistent with the Landau-Migdal p-h interaction isovector part are described in details in Ref. [15] where the parameters of the model for $^{208}$Pb are also given. These parameters (see Table 1 of Ref. [15]) are obtained from the description of the observed neutron and proton single-quasiparticle spectra in $^{208}$Pb. The values of $f^{ex} = -2.876$ and $f^{in} = 0.0875$ for the strength parameters of the isoscalar part of the Landau-Migdal p-h interaction, $F(r) = C(f^{ex} - (f^{ex} - f^{in})f_{WS}(r))$, where $C = 300$ MeV·fm$^3$, are determined from the condition that the isoscalar 1$^-$ spurious state, corresponding to the center-of-mass motion, should be at zero energy [16], and from the fit to the experimental value of the centroid energy of the ISGMR, $\omega_1 = (13.96 \pm 0.20)$ MeV [17]. The parameterization of the energy dependence of the imaginary part of the effective optical-model potential, $W(\omega)$, is given in Refs. [7,8,13]. The strength parameter $\alpha = 0.07$ MeV$^{-1}$ is found from the description within the model of the observed ISGMR $(FWHM)^{exp} = (2.88 \pm 0.20)$ MeV [17]. The strength of the real potential, $P(\omega)$, added to the mean field, is obtained from the imaginary part, $W(\omega)$, using the corresponding dispersive relationship [18]. We point out that all the model parameters have now been determined so that the main properties of any multipole isoscalar giant resonance in the $^{208}$Pb nucleus can be calculated without the need for additional parameters. This statement is naturally applied to the ISGMR2 as well.

In our analysis of properties of high-energy ISM excitations we first calculate the strength functions $S_{V_{0,i}}(\omega)$. In Figs. 1a and 1b we show the calculated fractions of the energy-weighted strength functions $y_i(\omega)$, obtained within the PHDOM for the external fields $V_{0,i}$, which are associated with the ISGMR ($i = 1$) and the ISGMR2 ($i = 2$), respectively. The parameter $\eta$ in the expression for $V_{0,2}$ (4) is evaluated in accordance with the method described in Sect. 2, obtaining the value of $\eta = 77$ fm$^2$. To demonstrate the contribution of the Landau damping to formation of the considered giant resonances, we show in Figs. 1a and 1b the fractions $y_i^{cRPA}(\omega)$ calculated within the continuum-RPA version, i.e. in the approximation $W(\omega) = P(\omega) = 0$. As follows from the results presented in Fig. 1a, the ISGMR in such heavy nucleus as $^{208}$Pb exhibits, taking the spreading effect into account, a well-formed resonance. The centroid energy $\omega_1 = 13.8$ MeV and total width $\Gamma_1 = 2.9$ MeV calculated within our model for the excitation energy interval 10-35 MeV, are in agreement with the above-mentioned experimental quantities [17]. Nonetheless, due to the Landau damping the fraction $y_1(\omega)$ shown in Fig. 1a exhibits a gross structure: along with the main maximum there is a small peak at the energy ~17 MeV. In contrast to the ISGMR, its overtone does not exhibit a well-formed resonance (Fig. 1b).

Considering the energy-averaged ISM double transition density, we show in Fig. 2, 3 and 4 the calculated transition density $\rho(r,r',\omega)$ of Eq. (1) for $r'=r$ at three energy points: in the vicinity of the maxima of the ISGMR and of the ISGMR2, and between these maxima. This transition density is compared with factorized projected and microscopically corrected classical double transition densities, $\rho_i(r,r,\omega) = (\rho_i(r,\omega))^2$ and $\rho_{c,i}^\Lambda(r,r,\omega) = (\rho_{c,i}^\Lambda(r,\omega))^2$, defined by Eqs. (14), (15) and (10)-(13), respectively. The comparison is given in Fig. 2 (i=1) and Fig. 3 (i=2). It is clearly seen from Fig. 2, that only near the maximum of the ISGMR the quantities $\rho(r,r,\omega)$, $\rho_{i=1}(r,r,\omega)$ and $\rho_{c,i=1}^\Lambda(r,r,\omega)$ are close (especially near the nuclear surface). For other energies these quantities are quite different. As follows from Fig. 3, near the maximum of the ISGMR2 the quantities $\rho(r,r,\omega)$, $\rho_{i=2}(r,r,\omega)$ and $\rho_{c,i=2}^\Lambda(r,r,\omega)$ are not so close. In Fig. 4 we plot results for i=1 at a vicinity of the ISGMR. As expected, the difference between the corresponding microscopic and classical transition densities appears at the giant-resonance tails.

In Figure 5, we plot the quantity $\rho(r,r',\omega)$ as a function of $r$ for the fixed value case of $r'$ and excitation energy $\omega$ in a vicinity of the ISGMR and ISGMR2, to trace the radial-dependence change from the one-node to two-node dependence.

We compare the projected and microscopically corrected classical transition densities, $\rho_i(r,\omega)$ and $\rho_{c,i}^\Lambda(r,\omega)$, at a vicinity of the ISGMR (i=1) (Fig. 6) and its overtone (i=2) (Fig. 7). As expected, the difference between the corresponding microscopic and classical transition densities is larger at the giant-resonance tails.

Further, we consider the energy-averaged strength function $S_{V_0,q}(\omega)$ of Eq. (19), which determines in the Born approximation the excitation cross-section of the ISGMR and its overtone in $^{208}$Pb by 240 MeV α-particle scattering. To avoid an appearance of small negative values of the strength function $S_{V_0,q}(\omega)$, we take the corresponding external field as $V_{0,q}(r) = \frac{\sin(qr)}{qr} - 1$. In such a way we restore a weak violation of the model unitarity [14]. It is natural that the strength function $S_{V_0,q}(\omega)$ (Figs. 8 and 9) exhibits the maxima, corresponding to the ISGMR and its overtone. In Figs. 8 and 9 we show the strength function evaluated via energy-averaged microscopic double transition density $\rho(r,r',\omega)$ in comparison with the strength functions evaluated with the use of the factorized projected and microscopically corrected classical transition densities, $\rho_i(r,\omega)$ and $\rho_{c,i}^\Lambda(r,\omega)$. As follows from this comparison, the description with the use of the factorized projected transition densities reproduces satisfactorily the "exact" description in a vicinity of the ISGMR (i=1) and ISGMR2 (i=2).

The quality of the approximate descriptions of the strength function $S_{V_0,q}(\omega)$ (Figs. 8 and 9) illustrates the ratios $Q_{V_0,i}(\omega)$ of Eq. (20) and $Q_{c,i}^\Lambda(\omega)$ of Eq. (21) shown in Figs. 10 (i=1) and 11 (i=2), respectively. As follows from consideration of these ratios, the approximate description of $S_{V_0,q}(\omega)$ in terms of the factorized projected double transition densities is preferable, as compared with the description in terms of the microscopically corrected classical double transition densities.

## IV. CONCLUDING REMARKS

We have presented in this work results of the first application of the recently developed particle-hole dispersive optical model (PHDOM) to calculate the energy-average double transition density for the isoscalar monopole excitations in $^{208}$Pb over a wide range energy region, which includes the energy regions of the ISGMR and its overtone (ISGMR2). The corresponding energy-averaged strength functions were also analyzed. The calculated double transition density was studied as functions of the excitation energy and position. The results were compared with the double transition densities constructed on projected and classical collective model transition densities for the ISGMR and ISGMR2. We have demonstrated, in particular, that in the intermediate excitation energy region (between the energies of the ISGMR and ISGMR2) the double transition density is quite different from that obtained from the classical transition densities, which are commonly used in the analysis of hadron inelastic scattering cross sections for the ISM excitations. As a first step towards to implications of the obtained results on the experimental analysis of hadron excitation cross sections, we have demonstrated possibilities of the model in employing the microscopic double and projected single transition densities for the description, within the Born approximation, of the inelastic α-scattering at zero angle cross sections for ISM excitations. Such a study is planned to be continued.

## ACKNOWLEGEMENTS

This work was supported by the Russian Foundation for Basic Research under grant No. 15-02-08007-a (M.L.G., B.A.T., M.H.U.), by the US Department of Energy under grant No. DE-FG02-93ER40773 (S.S.), and the Competitiveness Program of National Research Nuclear University «MEPhI» (M.H.U.). S.S is very grateful to the nice hospitality of the NRNU «MEPhI» during his visit and acknowledges the support by the Ministry of Education and Science of the Russian Federation for the Competitiveness Program of the NRNU «MEPhI», under grant No. 02.a03.21.0005. M.L.G. is very grateful to the nice hospitality of Cyclotron Institute during his visit supported by the Texas A&M University.

**FIGURE CAPTIONS:**

**Fig. 1.** Fractions of the energy-weighted strength functions $y_i(\omega)$ (the solid thick line) in comparison with $y_i^{cRPA}(\omega)$ (the solid thin line) for the ISGMR (i=1) and ISGMR2 (i=2) (Figs. 1a and 1b, respectively).

**Fig. 2.** The ISM double transition density $\rho(r,r,\omega)$ calculated at the different excitation energies: 13.8 MeV (a), 23 MeV (b), 33 MeV (c) (the solid thick lines) in a comparison with the factorized projected and microscopically corrected classical double transition densities, $\rho_i(r,r,\omega)$ and $\rho_{c,i}^{\Lambda}(r,r,\omega)$ (the solid thin and dashed lines, respectively), for i=1.

**Fig. 3.** Same as Fig. 2 but for i=2.

**Fig. 4.** Same as Fig. 2 for i=1, but at the different excitation energies in a vicinity of the ISGMR: 10.8 MeV (a), 13.8 MeV (b), 16.8 MeV (c).

**Fig. 5.** The ISM double transition density $\rho(r,r',\omega)$ calculated at the different excitation energies: 13.8 MeV (a), 23 MeV (b), 33 MeV (c) for the several fixed quantities $r'=5$ fm, $r'=7$ fm and $r'=9$ fm (the solid thick, solid thin and dashed lines, respectively).

**Fig. 6.** A comparison of the projected and microscopically corrected classical single transition densities, $\rho_i(r,\omega)$ (the solid thick lines) and $\rho_{c,i}^{\Lambda}(r,\omega)$ (the solid thin lines), $-\rho_{c,i}^{\Lambda}(r,\omega)$ (the dashed lines). The comparison is given for the excitation energies: 10.8 MeV (a), 13.8 MeV (b), 16.8 MeV (c) in a vicinity of the ISGMR (i=1).

**Fig. 7.** Same as Fig. 6 but for the excitation energies: 28 MeV (a), 33 MeV (b), 38 MeV (c) in a vicinity of the ISGMR2 main peak (i=2).

**Fig. 8.** The strength function $S_{V_{0,q}}(\omega)$ shown (the solid thick line) in a comparison with the strength function, corresponding to the same external field but calculated with the use of the projected double transition densities $\rho_i(r,r',\omega)$ (the solid thin red line) and microscopically corrected classical double transition densities $\rho_{c,i}^{\Lambda}(r,r',\omega)$ (the dashed blue line) for the ISGMR (i=1).

**Fig. 9.** Same as Fig. 8 but for the ISGMR2 (i=2).

**Fig. 10.** A comparison of the ratios of $Q_{V_0,i}(\omega)$ of Eq. (20) (the solid thick line) and $Q_{c,i}^{\Lambda}(\omega)$ of Eq. (21) (the solid thin line) for the ISGMR (i=1).

**Fig. 11.** Same as Fig. 10 but for the ISGMR2 (i=2).

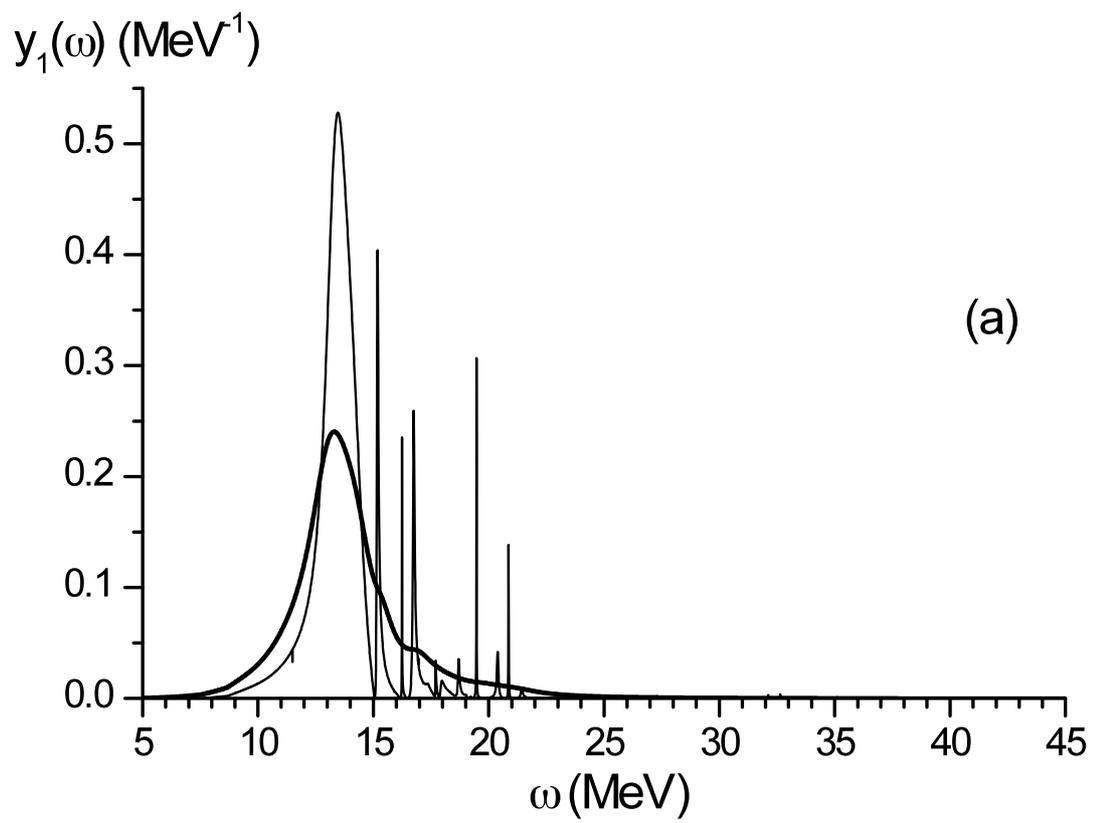

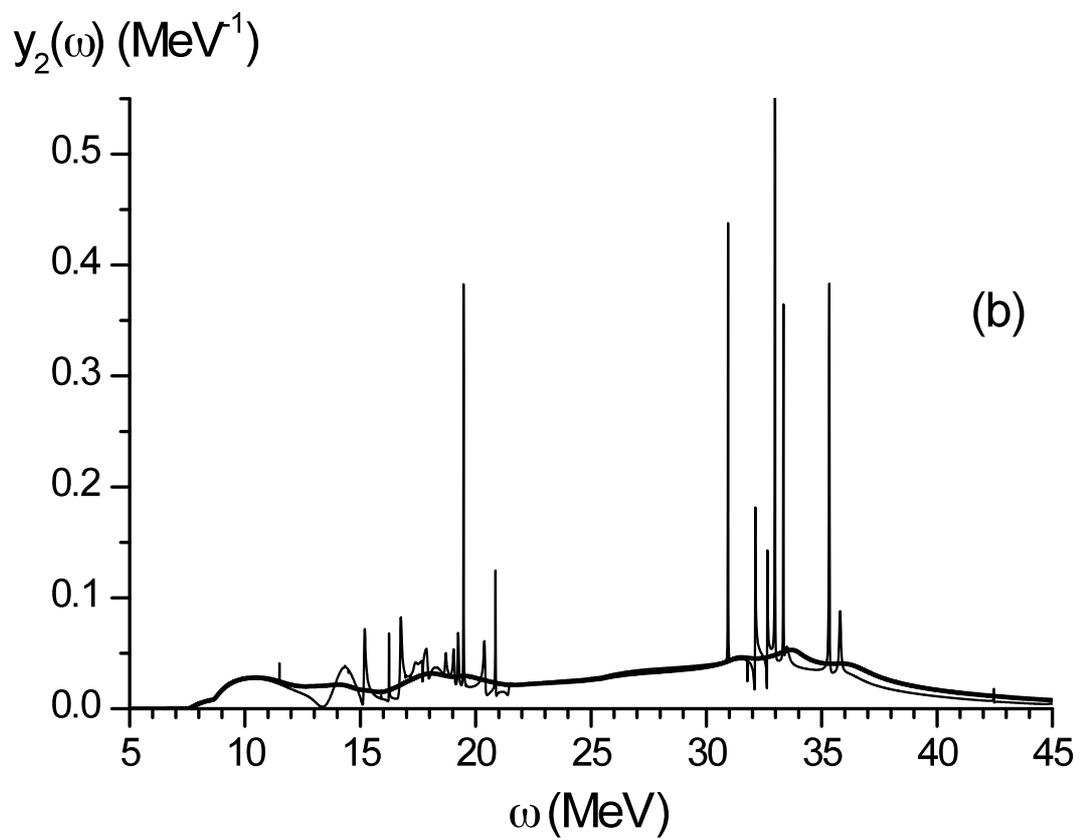

**Fig. 1**

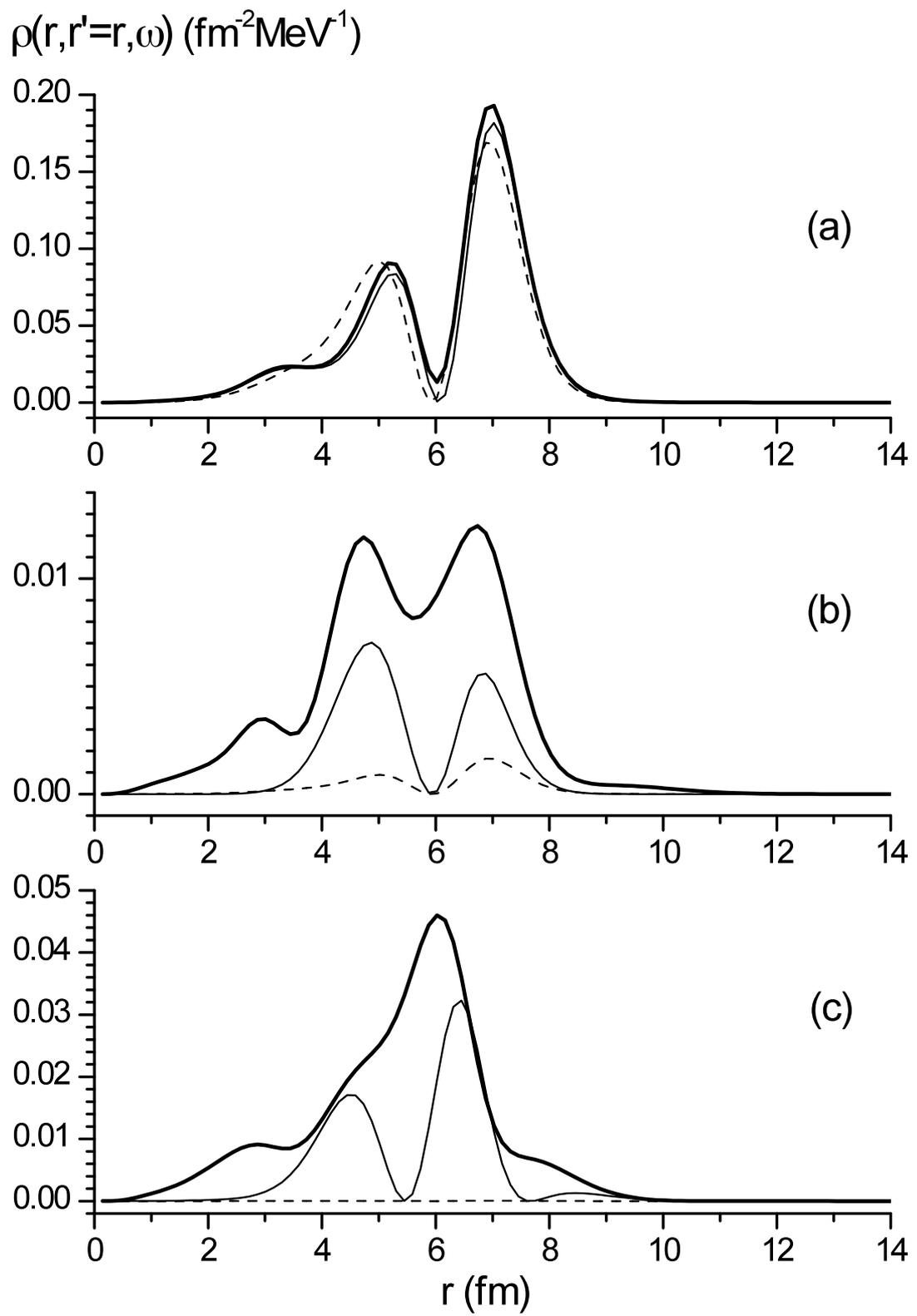

Fig. 2

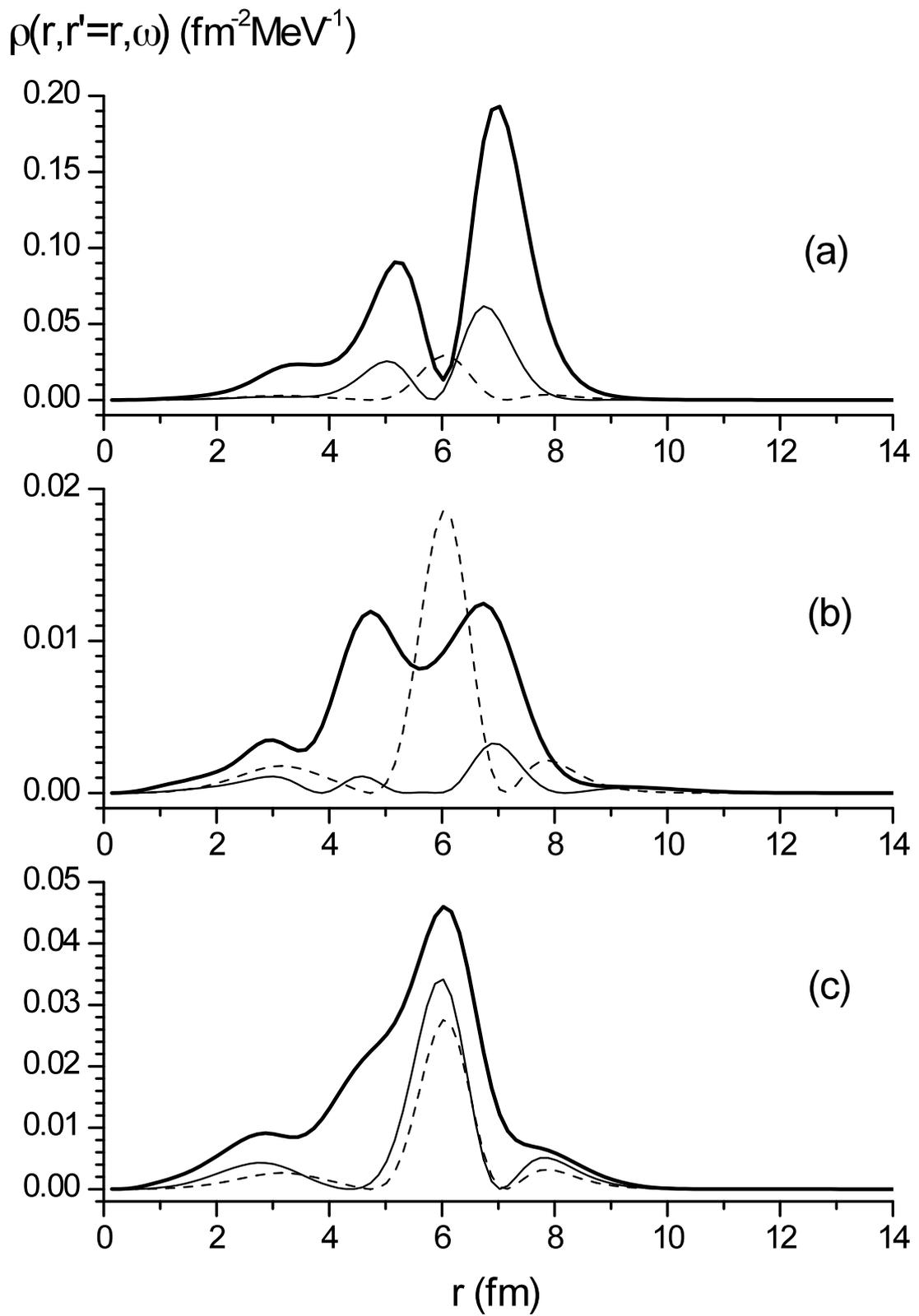

Fig. 3

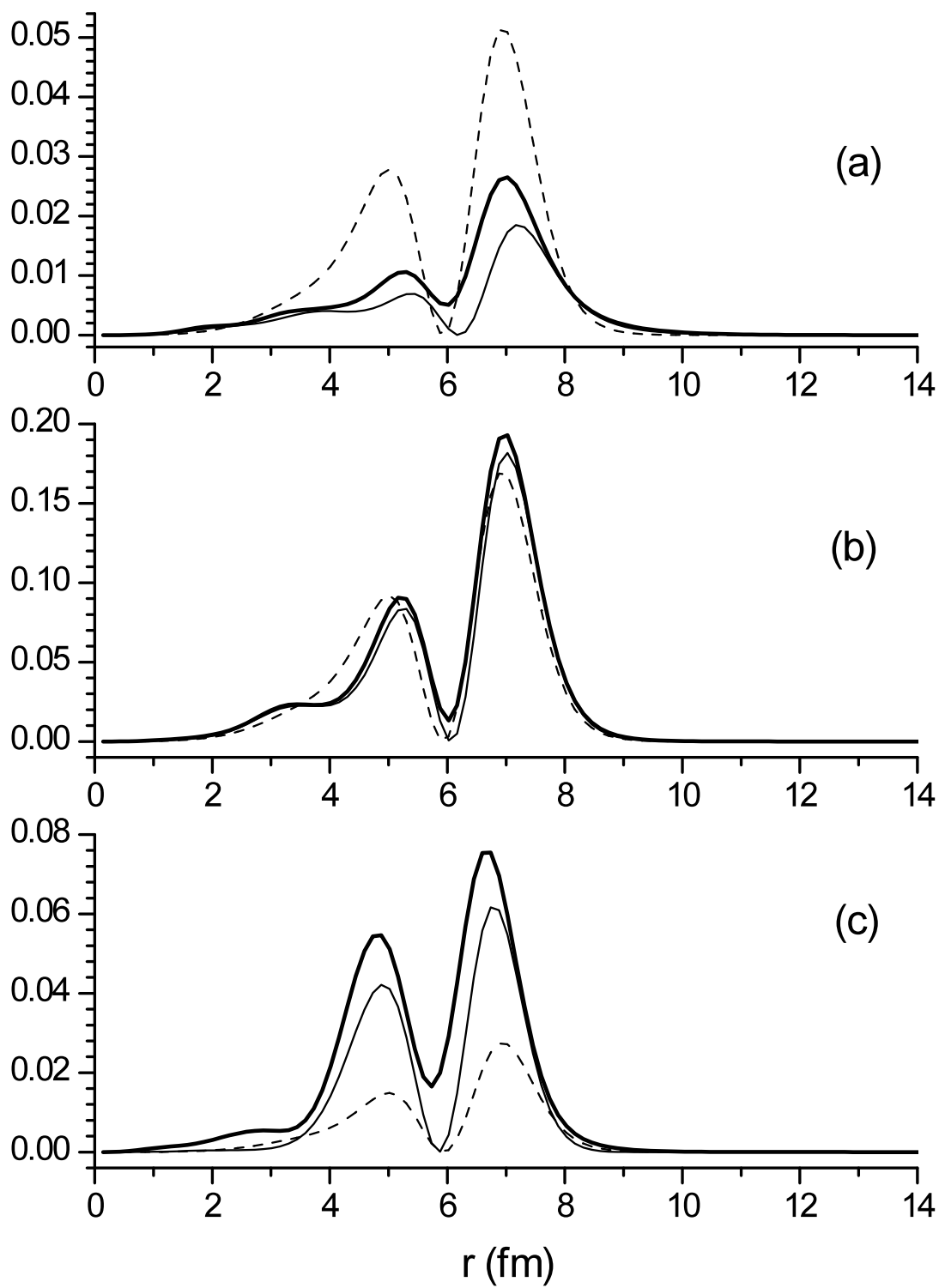

Fig. 4

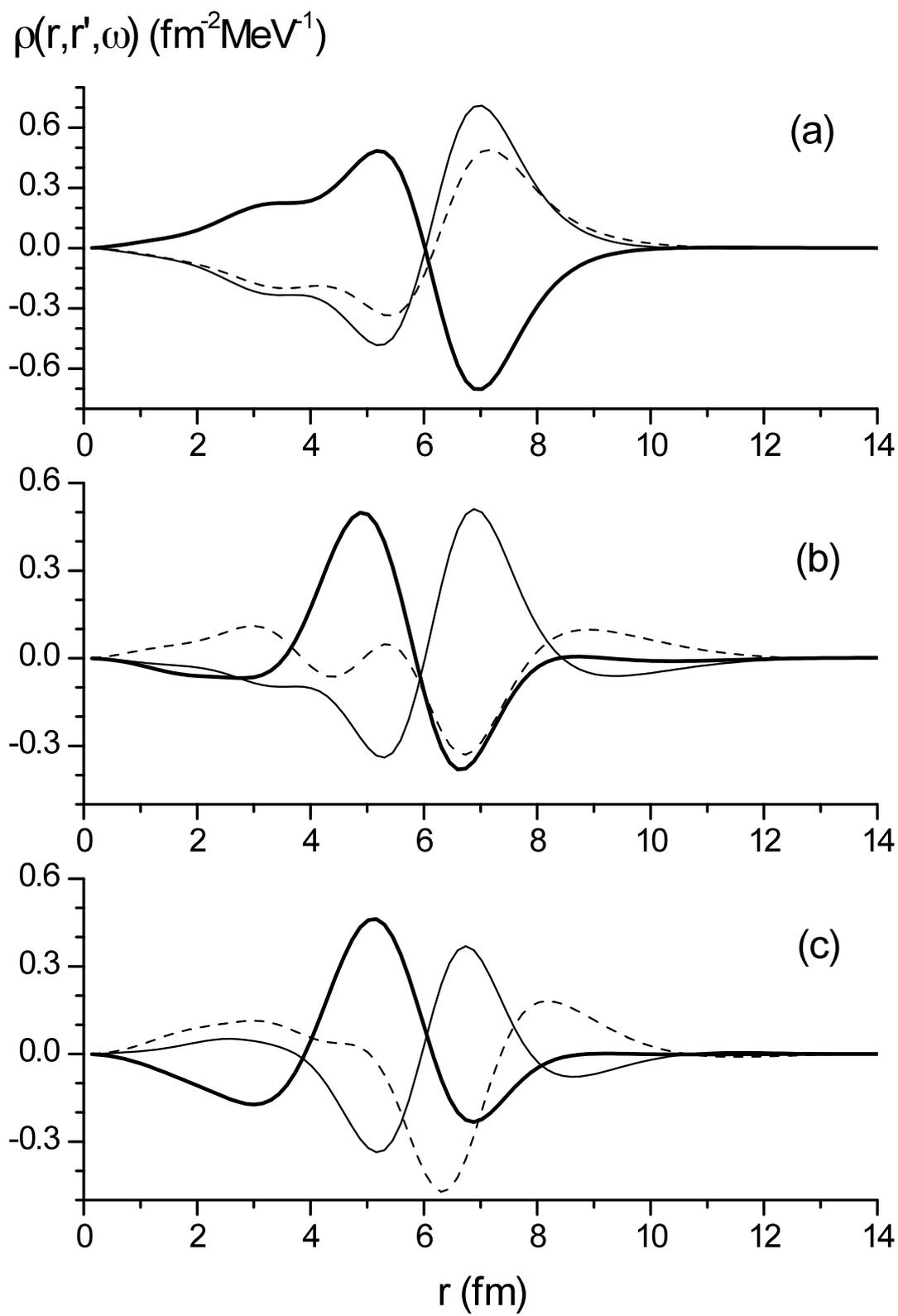

Fig. 5

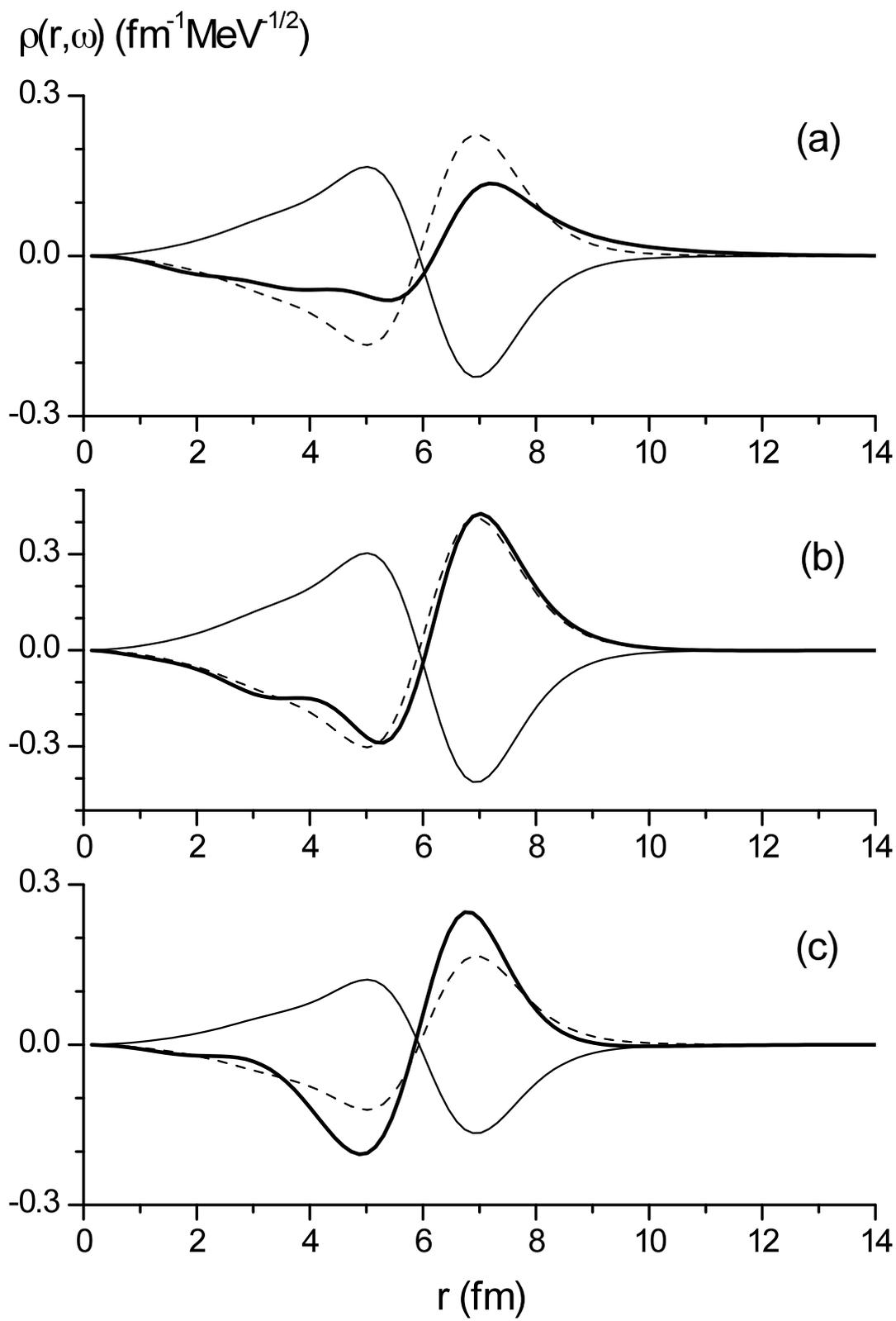

Fig. 6

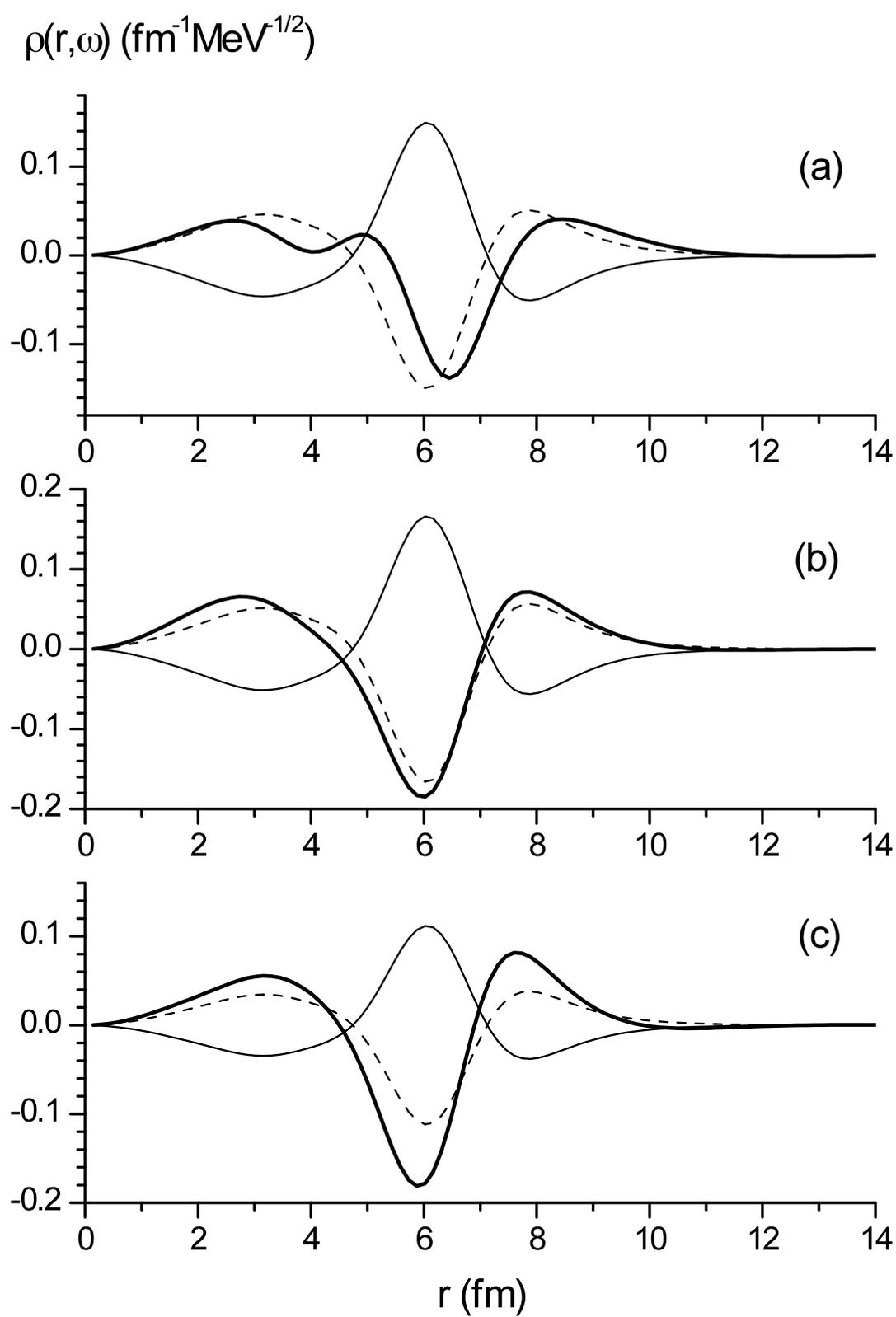

Fig. 7

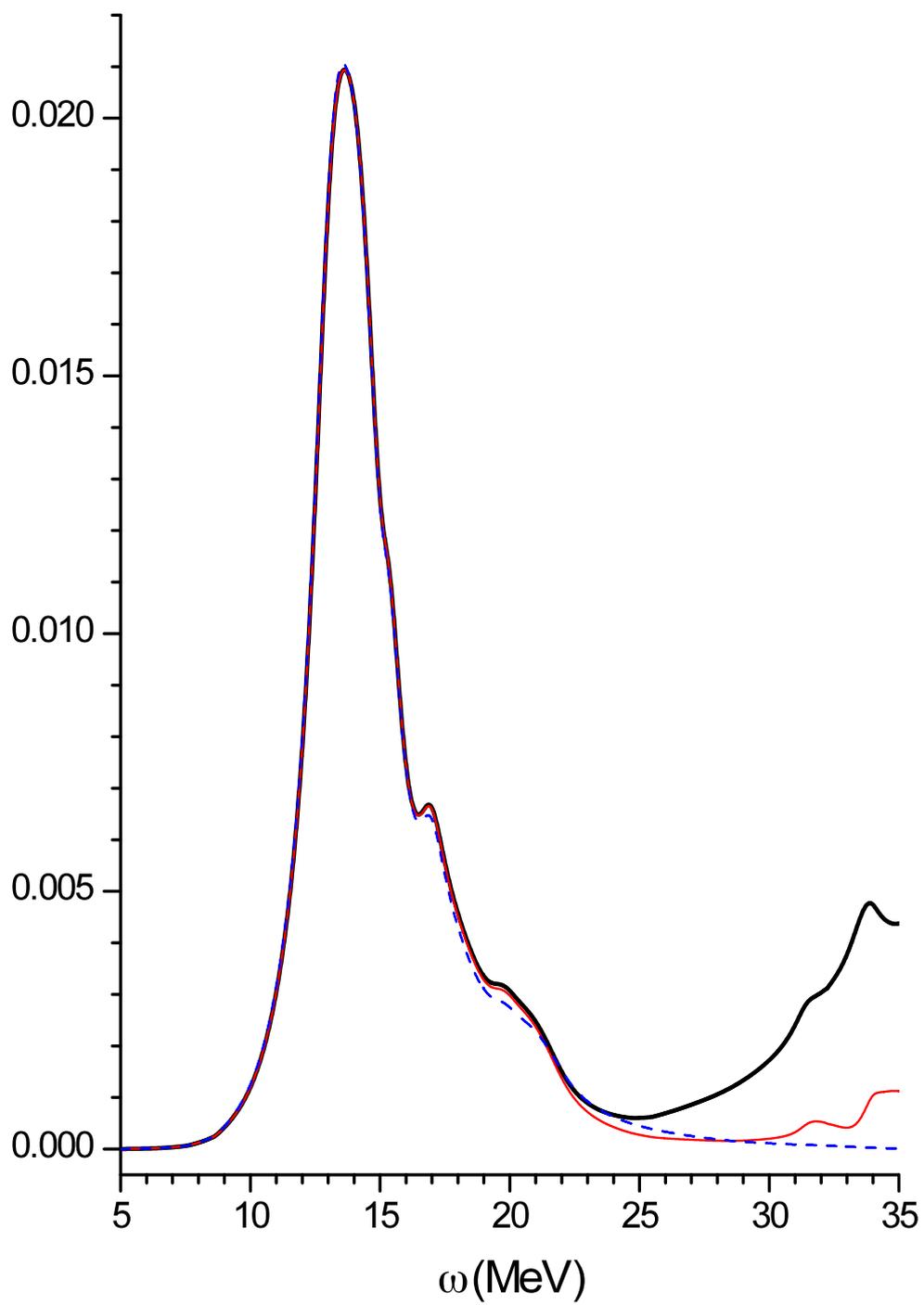

**Fig. 8**

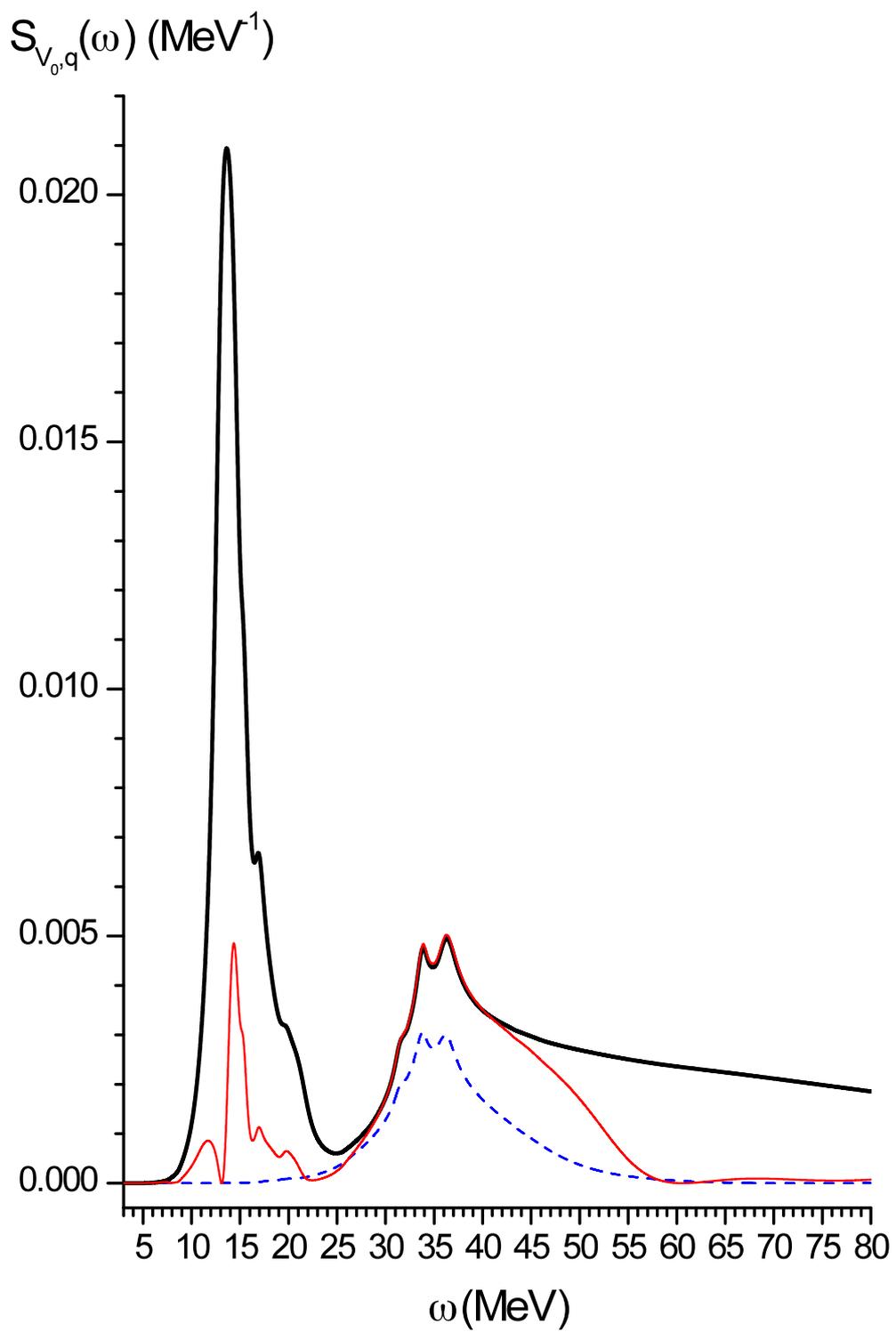

Fig. 9

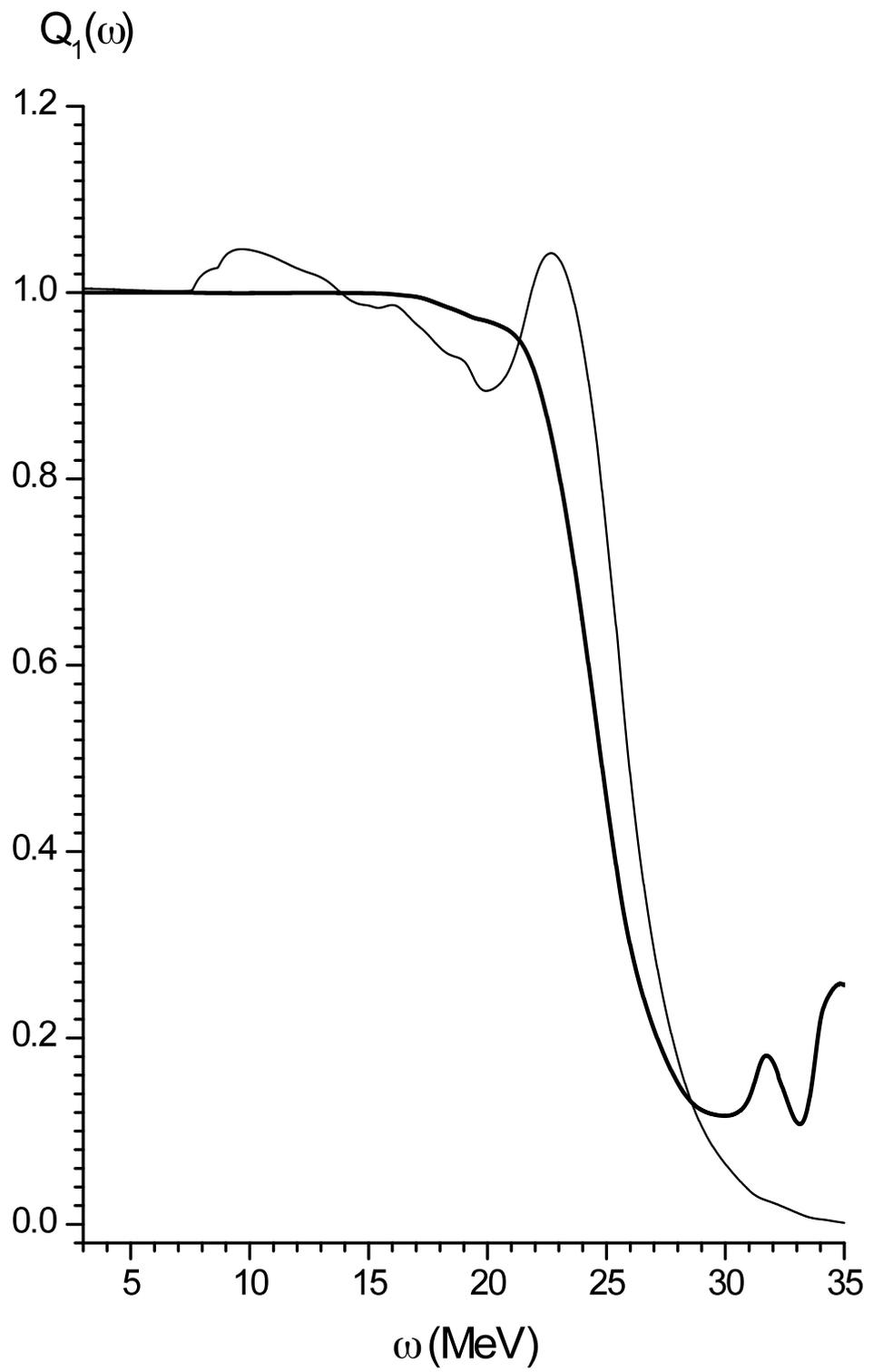

**Fig. 10**

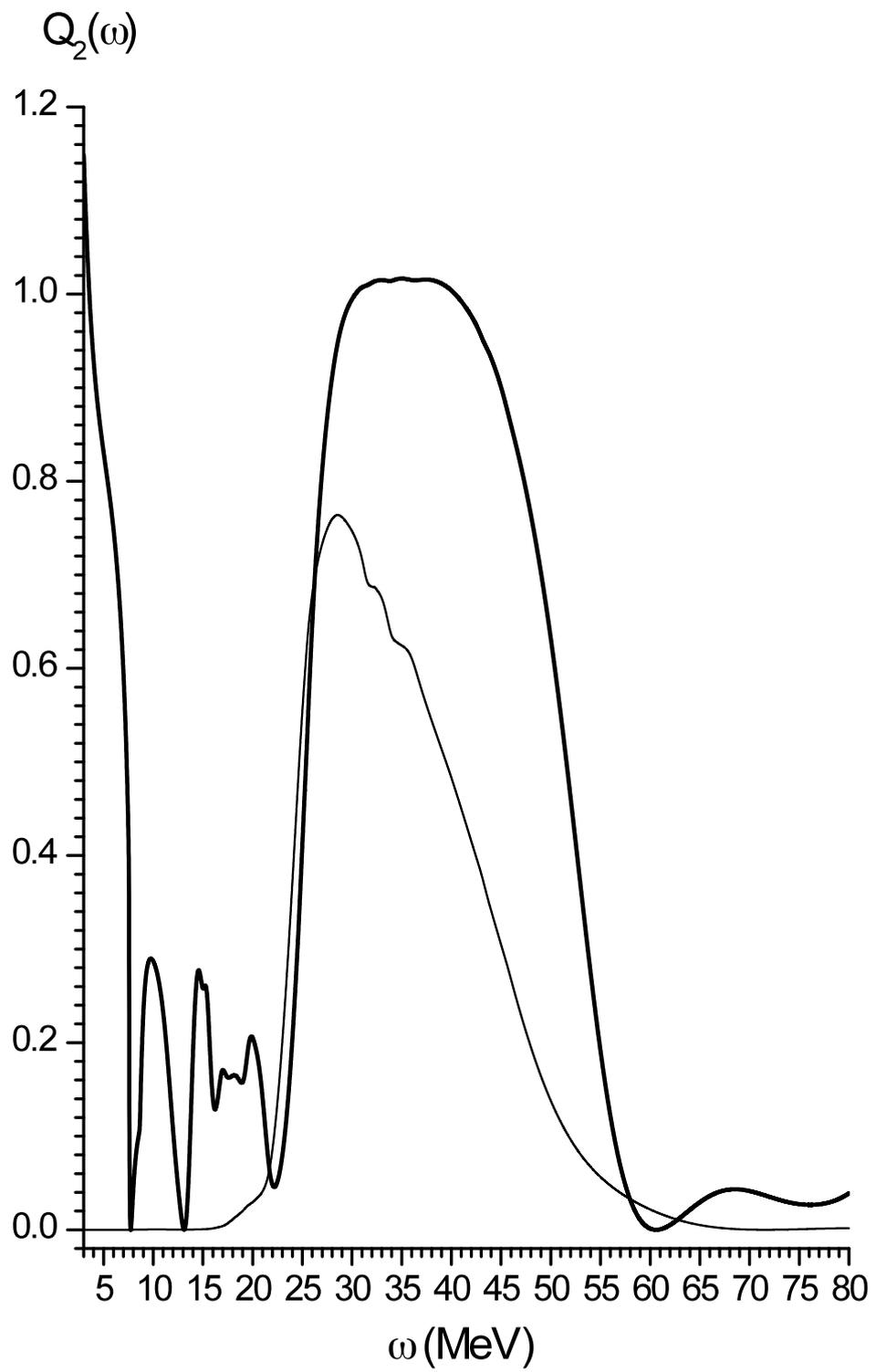

Fig. 11